\documentclass[12pt, a4paper]{article}

\usepackage{graphicx}
\usepackage{amsmath}

\title{Emergence of Hierarchy on a Network of \\ Complementary Agents}

\author{M. Copelli$^1$\thanks{Corresponding author:
mcopelli@if.uff.br}, R. M. Zorzenon dos Santos$^1$ and J. S. S\'a
Martins$^2$ \\ \small $^1$ Instituto de F\'\i sica, Universidade
Federal Fluminense \\ \small Av. Litor\^anea, s/n - Boa Viagem,
Niter\'oi, RJ, Brazil \\ \small $^2$ Colorado Center for Chaos and
Complexity/CIRES and Department of Physics \\ \small University of
Colorado, CB 216, Boulder, CO, 80309, USA }

\begin{document}

\maketitle 

\abstract{ Complementarity is one of the main features underlying the
interactions in biological and biochemical systems. Inspired by those
systems we propose a model for the dynamical evolution of a system
composed by agents that interact due to their complementary attributes
rather than their similarities.  Each agent is represented by a
bit-string and has an activity associated to it; the coupling among
complementary peers depends on their activity. The connectivity of the
system changes in time respecting the constraint of
complementarity. We observe the formation of a network of active
agents whose stability depends on the rate at which activity diffuses
in the system.  The model exhibits a non-equilibrium phase transition
between the ordered phase, where a stable network is generated, and a
disordered phase characterized by the absence of correlation among the
agents. The ordered phase exhibits multi-modal distributions of
connectivity and activity, indicating a hierarchy of interaction among
different populations characterized by different degrees of
activity. This model may be used to study the hierarchy observed in
social organizations as well as in business and other networks.  }


\section{Introduction}

In the last years great attention has been paid to the study of
different complex systems which share the common feature of organizing
themselves in networks. Examples include social
networks~\cite{Epstein96,Castellano2000}, the
Internet~\cite{Huberman99, Barabasi00WWW}, food
webs~\cite{Caldarelli98}, metabolic networks~\cite{Jeong00metabolic},
immune networks~\cite{Cohn2000,Zorzenon95, Zorzenon98}, economic
networks~\cite{Paczuski00,Zimmermann01} and ecological
networks~\cite{Lassig01}. Despite their differences, these systems may
be cast in a common framework regarding the structure of their
connectivity, being mostly classified as either regular,
scale-free~\cite{Barabasi99Sci}, small-world~\cite{Watts98} or random
networks. Scale-free networks are characterized by power law
distributions, indicating there is no typical number of connections
per site, whereas the small-world phenomenon is associated with high
clustering coefficient and distances between sites that increase only
logarithmically with system size~\cite{Watts98}, often presenting
exponential connectivity distributions~\cite{Barabasi99Sci}.

So far, the models proposed to study the formation and dynamics of
networks are mainly based on two approaches: dynamical processes
taking place in static network structures and the study of the
dynamics of the network structures themselves. Most of the
characterization of the network structures is focused on the dynamics
of aggregation of new nodes and links to a growing network. A key
point in the study of such networks is that their dynamics are based
on nodes which lack identity. Or, to put it correctly, their identity
and therefore their mutual affinity is uniquely determined by their
connectivity.  For instance, scale-free networks are obtained by
linking newly introduced nodes to old ones with a probability
proportional to the number of links of the latter~\cite{Barabasi99Sci,
Newman01d}. In other words, it is based on a continuous growth of
nodes with preferential attachment. When these complex networks are
used as a substrate to study a given dynamical process, the network
itself usually remains unchanged while the states associated to the
nodes evolve according to pre-determined rules~\cite{Watts98,
PastorSatorras01PRL, PastorSatorras01PRE, Bernardes01b}. A mixed case
has also been considered in the framework of game
theory~\cite{Zimmermann01}, where not only the state of the nodes
evolve in time, but also their connectivity. In the last case the
topological changes are biased by some state-dependent rules.



Although it has been pointed out that many of the network structures
mentioned above are scale-free, the assumed rich-gets-richer dynamics
that leads to this kind of structure may be appropriate only for some
cases~\cite{Barabasi00WWW}. It is not clear, for instance, whether
biological systems typically belong to this class of networks, since
the addition of new nodes does not necessarily occur continuously, and
certainly not without a strong dependence on the nodes identity. Many
biological systems organize themselves through pattern recognition
based on molecular interactions. Molecules, natural or synthetic, are
able to replicate or interact when their shapes and chemical features
are complementary. Depending on the way atoms or group of atoms are
distributed spatially in a molecule, it may fit into chemical ``nooks
and crannies'' of another~\cite{Rebek94}. Thus complementarity is
essential for pattern recognition interactions underlying biological
and biochemical processes.  It is the basis of the DNA replication
mechanisms and the generation of the immune responses, among many
other important biological processes. In the case of the immune
system, the control of foreign and harmful elements (antigens) to the
organism is based on the recognition between molecular receptors and
antigenic determinants through lock-and-key interactions. Another
example comes up when we think of symbiotic species in ecological
networks as complementary nodes, which generate a mutual reinforcement
when they interact. In this sense the role of complementarity can be
extended to social and economical systems, where agents need to
cooperate with other agents in order to prevail in a competing
environment.

Inspired on the complementary interactions observed in the immune
system and other biological and biochemical systems~\cite{Rebek94},
the model we present here may be considered as a third approach to
describe the formation of networks. The effective connectivity of the
system in this case emerges from the dynamical interaction among the
nodes, which is in turn inherently tied to their identity.  Using the
parlance of socio-economical models, each node is hereafter referred
to as an agent. In order for each agent to guarantee its participation
in the network, it needs not only to make use of its own capabilities,
but it also depends on the support of its complementary peers. The
maximum number of peers each agent may have is given by a neighborhood
which includes all possible complementary agents within some window of
tolerance. Thus, the maximum connectivity of each agent is fixed,
although large in most cases.  In principle all agents could be active
but in effect only a fraction of them remains activated due to
intrinsic regulation mechanisms. The intensity of the interactions
depends on the current state (hereafter referred to as ``activity'')
of the elements of the interacting pair, which will turn out to
dynamically mold their connectivity distribution. When a pair matches
due to the complementarity of their attributes, their activity will
increase, thus increasing the possibility of new interactions. Note
that this amounts to a mutually reinforcing mechanism which operates
on the activities of the nodes, not on their connectivity. The
appearance of new agents is introduced by diffusion of activity in
agent space, strengthening agents belonging to the similarity
neighborhood of the interacting agents. These new agents are chosen by
generating random mutations in one of the two interacting agents. The
newly added agents, however, will not survive unless they manage to
maintain themselves in the network, so that chance in this case is
subjected to regulation. As we shall see, the rate at which the
activity diffuses in the agent space is essential for the maintenance
and stability of the network.  Self-regulating mechanisms also take
place, preventing agents from increasing their activity too much.

Different from the approach used recently by Castellano {\it et
al.\/}~\cite{Castellano2000}, where the similarities between agents
play a key role in a social influence model, here we assume that
complementarity is the main mechanism in establishing the interactions
among agents. The question we address here is which kind of
organization is generated by individuals interacting locally by
complementarity, having a limited (although large) number of
interacting neighbors and whose connections are dynamically
established depending on their activity?

In the studies of social behavior it is very difficult to test
hypotheses concerning the relationship of individual behaviors and
macroscopic regularities observed. So far the main concern in social
sciences, especially in game theory and general equilibrium
theory~\cite{Epstein96}, has been focused on the stationary states
rather than the dynamical evolution of the system. Our aim in this
work is to study the essential micro or local mechanisms that lead to
the dynamical emergence of global regularities and properties in a
social community where individuals interact due to the complementarity
of their attributes.

Surprisingly, the fact that a given environment can support only
finite population, referred on the literature as the ecological
principle of carrying capacity, emerges naturally in our model. We
also obtain highly skewed distributions of activity. If we interpret
this activity as power, influence or wealth, the inequalities observed
in these distributions are characteristic of the actual human
society~\cite{Persky92}. Starting with agents having the same kind of
preferences and possibilities, but a unique neighborhood, the system
will end up with a multi-modal distribution of activities, since the
differences produced in their states after each interaction may be
amplified with time, generating the so called horizontal inequality
phenomenon~\cite{Epstein96}. 


\section{The model}
\label{sec:model}

The identity of each agent is defined by a bit string with length $b$,
each bit representing for instance the presence or absence of relevant
attributes. To each site $i$ ($i=1,...,2^b$) of this $b$-dimensional
hyper-cubic unit cell we associate a discrete variable $N_i \in
\{0,1,...,N_{max}\}$ which represents the {\em activity\/} of agent
$i$. If $N_i = 0$, we say that the $i$-th agent does not take part in
the network, or is inactive.

At each time step a parallel update of the individual activities of
the entire system is performed. The individual update of a given
activity $N_i$ follows two steps. In the first one, the activity of
agent $i$ is decreased with probability $p_i \equiv
\mbox{min}(1,N_i/N_{max})$.  This factor accounts for suppression
mechanisms generated by e.g.  competition for leadership or resources,
restricting the growth of an agent's activity, which is limited to a
maximal value of $N_{max}$:

\begin{equation} P\left(N^{\prime}_{i}|N_{i}\right) =
\binom{N_{i}}{N^{\prime}_{i}} p_i^{N_{i}-N^{\prime}_{i}}
(1-p_i)^{N^\prime_i}\; , \ \ \ \ \ \ N^{\prime}_i \leq N_i
\end{equation} where $N_{i}$ and $N^{\prime}_{i}$ are the current and
updated states of agent $i$, respectively, and
$P\left(N^{\prime}_{i}|N_{i}\right)$ is the probability with which the
activity of a site $i$ changes from $N_{i}$ to $N^{\prime}_{i}$ in a
time step. If the agents were allowed to be indefinitely activated,
they would simply accumulate activity and the system might never
attain a stationary state. In this sense the limitation on each agents
activity effectively amounts to the possibility of new agents being
incorporated into the network, according to the rules below.  In the
absence of interactions the agents activity will thus monotonically
decrease, on average, like $\left< N^{\prime}_{i}\right> = N_{i} (1 -
N_{i}/N_{max}) $. If $N_i$ eventually surpasses $N_{max}$, the
activity of the $i$-th agent is reset to zero.

The second step on the individual activity update accounts for the
mutually reinforcing interaction among agents, through which activity
increases~\cite{Arthur88}. We say that $j$ and $i$ are perfect
complementary matches if the Hamming distance $d(i,j)$ between them
(the number of different bits) is $b$ (in this case we write $j= \bar
i$).  However, an agent $i$ can interact not only with its perfect
complementary match $\bar i$, but also with agents that are partially
complementary. We define a parameter which measures this mismatch
window, so that an agent may interact with agents that have at most a
number Maxfit of equal attributes, or similarities.

If agents $i$ and $j$ are coupled, they will increase the total
activity by, at the most, $n_{ij}$ units, with a probability $p_{ij}$
per unit~\cite{Togashi01}, where

\begin{eqnarray}
n_{ij} & = & N_i N_j \\
p_{ij} & = & \frac{N_i N_j}{\binom{N_{total}}{2}}
\end{eqnarray}
and $N_{total}(t) = \sum_{i=0}^{2^b} N_i(t)$. The agent to which a
unit of activity is added will be similar to either $i$ or $j$: after
having chosen one of them randomly, each of its bits may be flipped,
with probability $M$ per bit (eq.~\ref{eq:mutation}). This provision
allows for a diffusion of the activity generated in agent space, which
will turn out to be essential for an agent's maintenance in the
network for two main reasons: 1) by diffusing their activity to a
similarity neighborhood, a given pair of agents strengthen other
(similar) agents with which they can further interact, thus creating a
supportive environment; 2) diffusion also helps protect an agent from
the suppression factor, preventing its activity from rising too fast
and reaching $N_{max}$. $M$ will thus be hereafter referred to as the
diffusion rate.

Denoting by $l_{k}$ the agent to which the $k$-th unit in the $(i,j)$
interaction will be potentially added, the dynamics can be described
by the iteration of equations~\ref{eq:mutation}-\ref{eq:increase}
below, for $k=1,\ldots,n_{ij}$ and $\forall i< j$ such that
$d(i,j)\geq \mbox{Maxfit}$:

\begin{eqnarray}
\label{eq:mutation}
P(l_{k}) & = & \frac{1}{2}M^{d(l_{k},i)}(1-M)^{b-d(l_{k},i)} +
\frac{1}{2}M^{d(l_{k},j)}(1-M)^{b-d(l_{k},j)} \\
\label{eq:increase}
P\left(N^{\prime\prime}_{l_{k}}|N^{\prime}_{l_{k}}\right) & = &
\delta\left(N^{\prime\prime}_{l_{k}},
N^{\prime}_{l_{k}}+1\right)p_{ij}+
\delta\left(N^{\prime\prime}_{l_{k}},
N^{\prime}_{l_{k}}\right)\left[1-p_{ij}\right]\; ,
\end{eqnarray}
where $\delta$ is the Kronecker delta. Note that the activity $i$ and
$j$ created remains within the pair with probability $(1-M)^b+M^b$.

Several quantities of interest are kept track of. As one would expect,
the stability of the activity of a given agent ultimately depends on
the activity of its complementary agents, including perfect and
slightly defective mismatches. To capture the instantaneous average
correlation between agents and their complements, we therefore
introduce a variable which measures the symmetry of the activation in
agent space:
\begin{equation} 
s(t) = \frac{2^b\sum_{i}N_i(t) N_{\bar{i}}(t) - \left(\sum_i
N_i(t)\right)^2} {2^b\sum_{i}N_i^2(t) - \left(\sum_i N_i(t)\right)^2}.
\end{equation} 
Note that $s=1$ when all $N_i = N_{\bar{i}}$.  We also compute the
fraction of agents with nonzero activity $f(t) = 2^{-b}\sum_{\{i |
N_i\neq 0\}}^{2^b} 1$ and the total activity $N_{total}(t)$ (see
above) of the network. The latter is very useful from the dynamical
point of view to help detect when the system reaches the stationary
state. 

It is instructive to observe how the populations of agents with low ($0 <
N_i/N_{max} < 10^{-2}$), intermediate ($10^{-2} \leq N_i/N_{max} < 10^{-1}$) and
large activity ($10^{-1} \leq N_i/N_{max} \leq 1$) change as the dynamics
evolve, which gives a first coarse-grained description of how activity is
allocated in the network. The fractions of active agents ($N_i \neq 0$)
falling into the three logarithmic bins above will be respectively called
$C_0(t)$, $C_1(t)$ and $C_2(t)$. Once the stationary state is reached, we
study in more detail the activity and connectivity distributions of the
system. We define an agent's local connectivity $k_i$ as the number of its
interacting peers which are active (i.e. we count the number of active
agents within the Maxfit-delimited region). Note that the maximal
connectivity of an agent

\begin{equation}
\label{eq:kmax}
k_{max} = \sum_{n=0}^{\mbox{Maxfit}} \binom{b}{n}
\end{equation}
is finite, but increases rapidly as the system size grows (e.g. $b=16$
yields $k_{max}=697$ for Maxfit=3 and $k_{max}=2517$ for
Maxfit=4). Therefore we have an interesting situation where the number
of {\em potential\/} connections is very large, yet much smaller than
the total number of agents (thus we are far away from mean field
interactions). Out of these $k_{max}$ potential individuals, an agent
will dynamically choose some, depending on how much activity the group
manages to gather.

In order to further characterize our analysis on the inequality of the
activity distribution in a given time step, we also calculate the Gini
coefficient, which is a well known index used in socio-economical
studies~\cite{Epstein96,Dragulescu}. The Gini coefficient measures the
inequality of the distributions by weighing the total activity (or
income, or power, etc) of the activated population by the activity of
the poorest elements. Compared to other measures of inequality, it has
the advantage of being applicable independently of the functional form
of the activity distribution (as opposed to the Pareto index, for
instance~\cite{Epstein96}). Agents are sorted by their activity and
the fraction $x$ of the least excited agents (i.e. the cumulative
population) is computed. Denoting by $y$ the cumulative activity
associated to those agents, the function $y(x)$ is called the Lorenz
curve. For perfectly ``egalitarian'' distributions ($N_i = N$,
$\forall i$), $y_{eg}(x)=x$. The Gini index is twice the area between
$y(x)$ and $y_{eq}(x)$, attaining zero for egalitarian distributions
and one if a single agent has all the activity in the system.

\section{Results}

For the sake of clarity in all the cases discussed below we have kept
Maxfit = 2 and $N_{max}=10^3$. In figures 1 and 2 we show the typical
behavior observed for single runs for $b=8$ and $b=12$, respectively.
We randomly select an initial fraction $f(0)$ of the total number of
possible agents and set their activities to $1\%$ of the maximal
activity.

\begin{figure}[!htb]
\begin{center}
\includegraphics[width=0.5\textwidth, angle=-90]{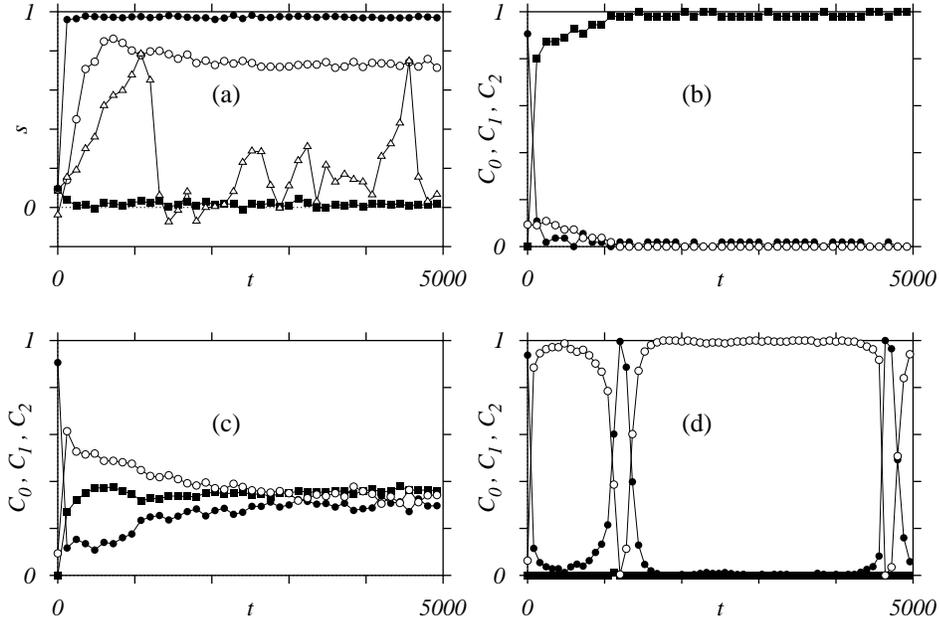}
\caption{\label{fig:b8} Typical dynamical behavior for $b=8$ with
$f(0)=0.25$ ($t$ in Monte Carlo steps): (a) the correlation $s$ among
agents for $M=0$ (squares), $10^{-4}$ (open circles), $10^{-2}$ (black
circles) and $10^{-1}$ (triangles). The fraction of active agents with
low (open circles), intermediate (black circles) and large (squares)
activity are shown from (b) to (d) for $M=0$, $10^{-4}$ and $10^{-1}$,
respectively.}
\end{center}
\end{figure}

The dynamics exhibited for $b=8$, although interesting in its own and
revealing the appearance of emergent properties, is slightly different
from the dynamics observed for larger system sizes (see the results
for $b=12$ below). Figure~\ref{fig:b8}(a) shows the behavior of the
correlation $s(t)$ among complementary agents for different values of
the diffusion rate $M$. For M = 0 there is a very low correlation
among complementary agents, since $s$ is essentially established by
the initial configuration. The activation will basically last among
these initially activated agents, due to the lack of diffusion of
activity in agent space. Increasing the diffusion rate $M$ from
$10^{-4}$ to $10^{-1}$, we observe that the correlation initially
increases, reaching a maximum value around one for $M=10^{-2}$, in
which case we interpret that a network is mounted. For $M=10^{-1}$ the
correlation oscillates non-periodically, indicating that a large rate
of diffusion does not guarantee the stability of the network, which
mounts for a while but does not last long. This last kind of behavior
is reminiscent of the ``global dynamical cascades'' observed in a
model of heterogeneous agents for economics~\cite{Zimmermann01}.

Figures~\ref{fig:b8}(b)-(d) show the result for the fraction of active
agents with low, intermediate and large activity for $M=0, 10^{-4}$
and $10^{-1}$ respectively. In the first case [Fig.~\ref{fig:b8}(b)],
without diffusion of activity in agent space, the system does indeed
reach a stationary state, with the overwhelming majority of the active
agents with large activity. The fact that no intermediate- or
low-activity agents are present turns out to be an indication of
failure ($s\sim 0$) of the network to self-organize hierarchically
(note that agents with $N_i = 0$ do not enter the statistics of the
$\{C_i\}$).  A mild amount of diffusion ($M=10^{-4}$) is enough to
help mounting a hierarchical structure, as shown in Fig.1(c). The
behavior is now completely different from the previous case and we do
observe the formation of a network. This is not only characterized by
the large value of $s$, but also by the stratified concentrations of
active agents, which stabilize in the long run. The stationary values
suggest a rough balance between low, intermediate and large activity
in order to sustain the structure of the network. This pattern
persists up to values $ M = {\cal O}(10^{-2})$. The results for
$M=10^{-2}$ (not shown) indicate the formation of a stronger and
highly activated network, with the majority ($\sim 80\%$) of the
agents with intermediate activity, and $\sim 20\%$ with large
activity, showing some hierarchy, where stronger agents are kept by a
large number of intermediate-activity agents and essentially none with
low activity.

For larger values of the diffusion rate, the formation of the network
is destabilized, as shown for $M=10^{-1}$ in Figure~1(d). We observe
oscillations in the values of $C_0(t)$ and $C_1(t)$, suggesting that
the network sustains itself for a brief period. However, this balance
is quickly destroyed by the rate at which activity disperses in the
network, which prevents the stronger agents and their complementary
partners to satisfactorily keep their dominance. The peak in $C_1$
[$t\sim 1100$ in Fig.~\ref{fig:b8}(d)] indicates that the activity
which the interacting agents (and their neighbors) create spreads too
much and too fast in agent space, eventually destroying their
prevalence and returning the network to a non-hierarchical situation,
where no particular pair or group of agents dominates the game. In the
following it may happen that a new dominant group emerges out of the
inherent fluctuations of the model, only to be destabilized later, in
a quasi-periodic fashion. This behavior is very robust and has been
exhaustively checked over many long runs for different initial
configurations. For higher diffusion rates ($M\simeq 0.4$), the peaks
described in the previous case disappear and $s$ fluctuates around
zero, indicating the complete absence of correlation among the
agents. In this case $C_0$ is essentially one, showing that the system
consists of a very large number of weakly activated (and disconnected)
agents. It is interesting to point out that in the whole range
$10^{-4} < M < 10^{-1}$ (where the network is mounted), the fraction
of active agents $f$ remains above $\sim 80\%$ in the stationary
state, indicating that the majority of agents is active.

\begin{figure}[!htb] 
\begin{center} 
\includegraphics[width=0.5\textwidth,angle=-90]{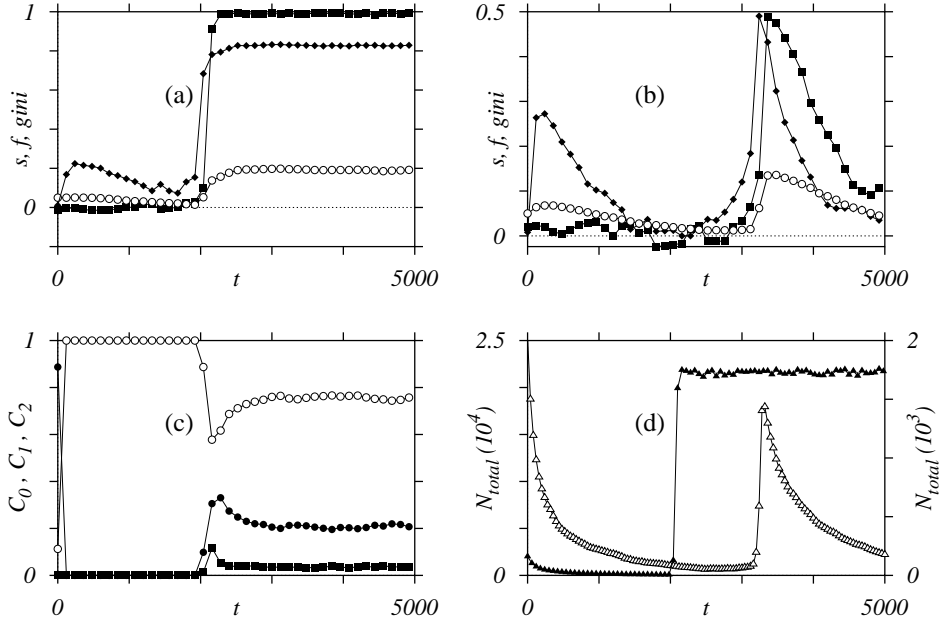} 
\caption{\label{fig:b12}Typical dynamical behavior for $b=12$,
starting with $f(0) = 0.05$ ($t$ in Monte Carlo steps): correlation
$s$ among agents (squares), fraction of active agents $f$ (open
circles) and Gini coefficient (diamonds) for $M=10^{-3}$ (a) and
$M=10^{-1}$ (b); fraction of active agents with low (open circles),
intermediate (black circles) and large activity (squares) for
$M=10^{-3}$ (c); and the total activity of the system (d) for
$M=10^{-3}$ (black triangles --- left axis) and $M=10^{-1}$ (open
triangles --- right axis).}
\end{center}
\end{figure}

Results for $b=12$ are shown in Figure 2 for two representative values
of the diffusion rate. The first interesting feature we observe when
$M=10^{-3}$ [Fig.~\ref{fig:b12}(a)] is that, unlike the 8-bit case,
there is a long transient period before the fraction of active agents
$f$ and the correlation $s$ stabilize. After that, the emergence and
establishment of the network then occur in a relatively short
time. Moreover, $f$ remains around $0.2$ in the plateau, signaling
that {\em the majority\/} of the agents have zero activity and do not
take part in the network. In the evolution of concentrations $\{C_i\}$
shown in Fig~\ref{fig:b12}(c), most of the active agents have low
activity, followed by agents with intermediate and large activity. The
same behavior is observed for different samples, although the
stationary concentrations may fluctuate around slightly different mean
values.  The larger 12-bit system is therefore much less excited and
sparser than the smaller 8-bit one: not only is the fraction of active
agents smaller, but also the activity of those agents is typically
less than for the smaller system.  Fig.~\ref{fig:b12}(b) shows the
results for a higher diffusion rate ($M=10^{-1}$) where again we
observe a long transient, after which a peak in $s$ and in $f$ occurs,
corresponding to an attempt to mount the network. The attempt fails
due to the instability generated by the high rate at which activity
diffuses. The system will keep spiking indefinitely, similarly to what
has been observed for 8 bits in the same parameter region. Finally, a
complementary view of these phenomena can be composed by the results
in Fig.~\ref{fig:b12}(d), which shows the evolution of the total
activity for both diffusion rates. For $M=10^{-3}$ the stabilization
of $s$ and $f$ is associated with the stabilization of $N_{total}$ at
a large value, while for $M=10^{-1}$, the peaks in $s$, $f$ and
$N_{total}$ occur simultaneously, the latter attaining a much lower
level (note the different vertical scales) than in the $M=10^{-3}$
case.

Finally, it is interesting to observe the behavior of the Gini
coefficient for both diffusion rates. Figures.~\ref{fig:b12}(a)-(b)
show that it departs from zero (since the initial agents have the same
activity) and immediately grows, indicating the immediate creation of
inequalities due to the fluctuations. Note however that during this
growth $s$ remains near zero, showing that the initial fluctuations do
not immediately lead to a structuring in agent space. The inequality
decreases a little and then increases, now lifting with it the
correlation. The Gini coefficient then stabilizes around $0.80$,
indicating a highly uneven distribution. 

\begin{figure}[!htb]
\begin{center}
\includegraphics[width=0.5\textwidth, angle=270]{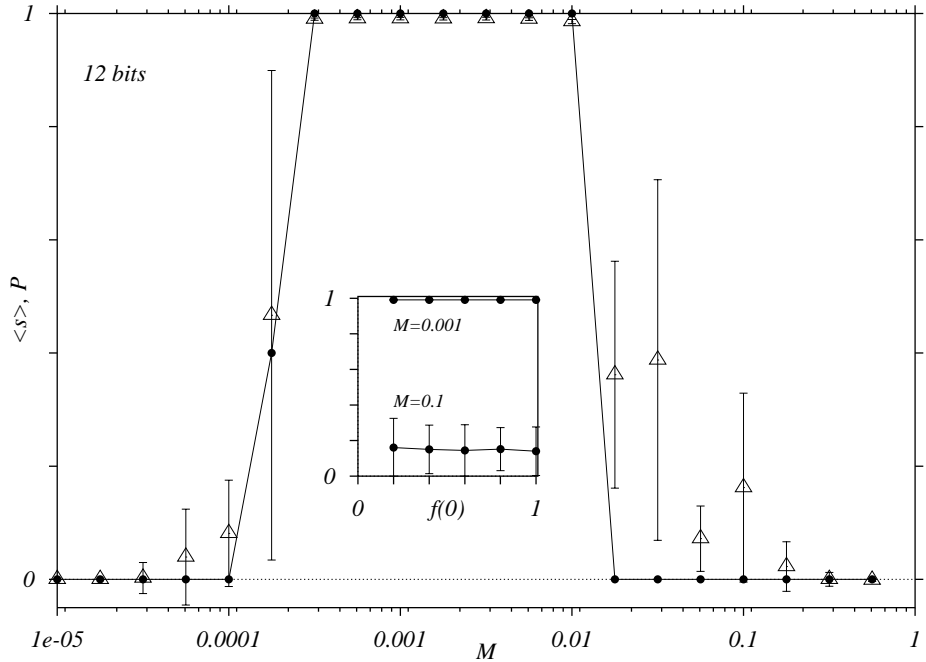}
\caption{\label{fig:bell} Time-averaged correlation $\left< s \right>$
(triangles) and ``mount probability'' $P$ (black circles --- see text
for details) as a function of $M$ for $12$ bits.  Inset: $\left<
s\right>$ as a function of the initial concentration for $M = 10^{-3}$
(upper curve) and $10^{-1}$ (lower curve). The results correspond to an
average over 50 runs.}
\end{center}
\end{figure}

The diffusion rate thus plays an extremely important role in
connecting the agents, its effect being: too low or too high diffusion
rates deteriorate the stability of the network, while intermediate
values increase the diversity of active agents in the right amount to
allow new pairs to match, thus providing the mechanism for the system
to adapt, incorporating new agents. Note that in the current model
this is the only source providing new active agents. The parameter
Maxfit plays a complementary role to that of the diffusion rate on the
regulation of the network. A large diffusion rate may eventually be
compensated by a smaller Maxfit, which renders the agents more
specific in the interaction with their mirrors (in the sense that it
reduces the number of interacting peers). However, Maxfit cannot be
too low lest there be not enough agents to create a supportive
neighborhood, specially for low diffusion rates.

The dependence of the formation of the network for $b=12$ on the
diffusion rate is summarized in Fig.~\ref{fig:bell}, which shows the
stationary value of $s$, averaged over 500 time steps and 50 runs, as
a function of $M$. In order to obtain this plot we have adopted the
following protocol: we consider that the network is mounted if $s$
stays larger than a threshold value $s_{min}=0.8$ (estimated from many
individual runs for the parameters under consideration) during a
period of 1000 time steps. This time interval is long enough for an
eventual peak not to be erroneously detected as a successful
attempt. The ``mount probability'' is defined as the fraction of times
that this criterion is satisfied. We observe then two transitions, one
at $M\simeq 10^{-4}$ and another at $M\simeq 0.03$. The large error
bars in $s$ around the transition points are the signature of the
peaks observed in Fig.~\ref{fig:b12}.  The first transition is mainly
provoked by the suppression factor: since the rate at which activity
disperses is very low, the initially correlated agents reinforce
themselves, leading to a fast increase in their activity which
strongly submits them to the suppression factor. In this case no
correlation is maintained long enough to mount the network. The second
non-equilibrium phase transition is due to the dynamics of the system
as discussed below.

Another interesting feature that emerges from this model is the weak
dependence on the initial conditions. For several values of $M$ we
have tested the evolution of the system for different initial
fractions of nonzero agents. What we have observed, for both 8 and 12
bits, is that the results are qualitatively the same for whatever
initial conditions we choose, except of course for the limiting case
$f(0)\simeq 0$, in which the activities eventually decay to zero due
to a lack of matching pairs. This result is clearly observed in the
inset of Fig.~\ref{fig:bell}, which shows the stationary value of $s$
as a function of $f(0)$ for 12 bits. The mechanisms behind this
independence can be understood on the basis of the transients of both
$N_{total}$ and $s$ (Fig.~\ref{fig:b12}). In order for the network to
stabilize, the unpaired initial agents will first be ``cleaned up'',
leaving mostly pairs that match within the margins set by Maxfit. The
dynamics automatically takes care of this initial dismantling, since
all those agents which do not have their matching pairs will gradually
decrease their activity [see Fig.~\ref{fig:b12}(d)]. This in turn
increases the probability that the remaining matching agents interact,
and stationarity is then achieved when activity increases and losses
are balanced.  This explains why the model is not very sensitive to
the initial conditions. Larger initial values of $f$ only increase the
time spent in the transient, therefore we have typically used
$f(0)=0.05$ for 12 bits. In the 8-bit case, the initial transient is
much shorter and of a different nature because the network somehow
manages to accommodate a much larger fraction of agents in highly
excited states, showing a different dynamics. Due to the small system
size all potential pairs will be effectively connected, which is not
the case for larger system sizes. This particular behavior might be
related to what is observed in population models and commonly referred
to as finite population effect~\cite{Sousa00}.

It is also important to understand how the behavior of the system
changes when the size of the space gets larger. Representative
simulations for $b=16$ (using $f(0)=0.015$) show that the fraction of
active agents $f$ in the stationary state is even smaller than for 12
bits, but the behavior of the concentrations $\{C_i\}$ remains
qualitatively the same, indicating that those are the relevant
quantities for larger system sizes. The dynamical behavior for 16 bits
exhibits the same features as for 12 bits, the corresponding plots
(not shown) being very similar to Fig.~\ref{fig:b12}. Even though a
more detailed analysis of the parameter space is currently under
investigation, it is clear that the effects of the diffusion rate $M$
follow the same trends both for larger systems and for smaller
systems: if it is either too large or too small, no balance can be
achieved and the agents will fail to establish a connected network. It
is interesting to note that for Maxfit=2 the absolute number of active
agents in the stationary state remains ${\cal O}(10^3)$ when one goes
from 12 to 16 bits, explaining the decrease in $f$. This number grows
when Maxfit=3, but still corresponds to a small fraction of the $2^b$
potential agents, for large systems, and it is the diffusion rate that
guarantees its stability. Without any direct suppression mechanism
acting on the number of agents belonging to the network, the local
rules of the model therefore lead to an effective ``carrying
capacity'' of the system~\cite{Epstein96}. The proper scaling of
Maxfit with the growth of the system is yet to be fully understood,
i.e., how does the stationary value of $f$ depend on the relation
between Maxfit and b?  We also conjecture whether a larger agent space
would allow the system to adapt more easily to perturbations, which
normally would destabilize the network for small systems.  These and
other questions concerning the robustness of the network on the
parameter space are currently under investigation and will be
published elsewhere.

\begin{figure}[!htb]
\begin{center}
\includegraphics[width=0.5\textwidth, angle=-90]{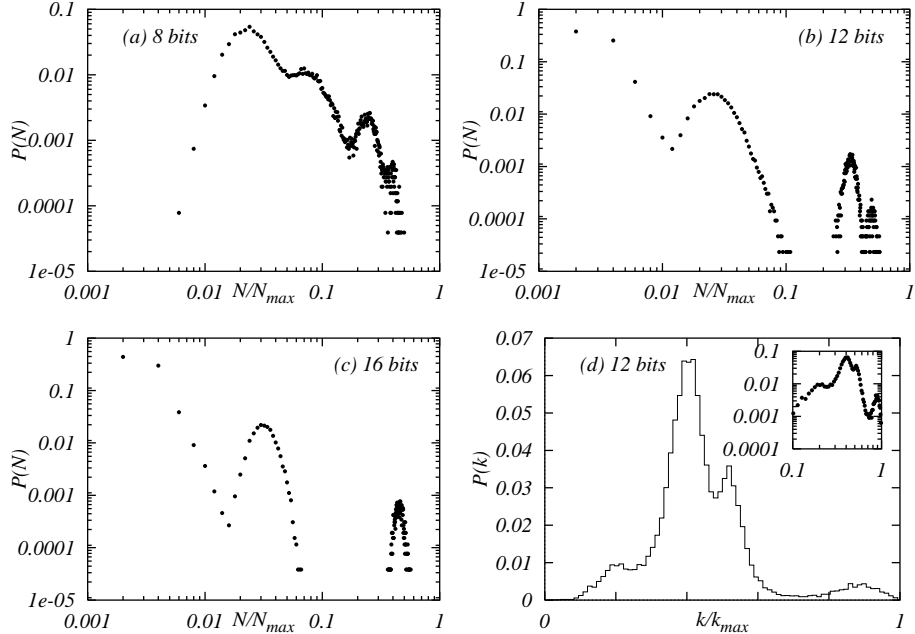}
\caption{\label{fig:histo}Activity distributions for $M=10^{-3}$:
(a) $b=8$, (b) $b=12$ and (c) $b=16$; connectivity distribution
for $b=12$ (d). Inset: same as (d) in log-log scale.}
\end{center}
\end{figure}

We finally turn to the characterization of the ordered phase, focusing
at intermediate values of $M$. Our analysis are based on the emergent
distribution of activity and connectivity in the network, which turns
out to be highly non-trivial. $P(N)$ measures how activity is
allocated, whereas $P(k)$ is the connectivity distribution.  These
distributions are calculated in the stationary state for $b=8$
($M=10^{-2}$) and for $b=12$ and $b=16$ ($M=10^{-3}$) over 30
runs. Results are shown in Fig.~\ref{fig:histo}. For the three cases
[Figs.~\ref{fig:histo}(a)-(c)], we obtain a multi-modal distribution
of activity. Therefore, the system does not have a single typical
activity, nor does it possess a scale-free activity distribution: as
opposed to that, we observe that there are four well-defined
``typical'' values of activity (other than zero) for 8 and 12 bits,
and three for 16 bits, defining a hierarchy in agent space. We
therefore justify in hindsight the arbitrary logarithmic binning as
defined by the variables $C_0$, $C_1$ and $C_2$, which are actually
reasonable coarse grained descriptions of the peaks of
Fig.~\ref{fig:histo}. It is also worth noting that, for 12 and 16
bits, there is a gap in the activity distribution, a feature that has
also been observed in larger system sizes.

The distribution presented in Fig.~\ref{fig:histo}(d) exhibits an even
richer behavior for 12 bits. Not only do we observe multi-modality
again, but also a non-monotonic behavior of the peaks. In other words,
the majority of the agents are ``connected'' to $\sim 40\%$ or $\sim
50\%$ of their potential collaborators, while two smaller groups
either have $\sim 20\%$ or $\sim 90\%$ of their possible connections
actually established. The distribution for 16 bits has similar
features, while for 8 bits $P(k)$ in its most ordered regime is simply
given by $P(k) = \delta(k,k_{max})$ (where $k_{max}$ is defined in
eq.~\ref{eq:kmax}), reflecting the fact that $f$ saturates at one.
The increase of the variance of the activity distribution corresponds
to an increase of the horizontal inequality. It depends on the rate at
which new agents are activated (diffusion of activity in agent space),
since new agents take some time until their activity is brought into
line with those already prevailing in the system.

\section{Conclusions}    

In this paper we proposed a network model of mutually reinforcing
agents by using a bit-string approach. The model is inspired in the
complementary interactions observed in biological and biochemical
systems. The interaction among the agents is based on their
identity. All agents have potentially the same connectivity although
the actual individual connectivities are established by dynamical
rules based on interactions that allow perfect and slightly defective
mismatches between agents. A suppression mechanism acting upon agents
limits the maximum activity that may be associated to each agent. New
agents may be added to the system through the interactions among
active agents, which allow the diffusion of activity in agent
space. The rich dynamical behavior exhibited by this kind of network
depends on the diffusion rate, the size of the mismatch window and the
maximum activity allowed to each agent.  The agents prospering in
activity may activate new agents.  For very low or very large
diffusion rates there is a lack of correlation between complementary
agents and the network is not formed, while for intermediate values of
this parameter we observe the formation of a stable network.

For small systems (8 bits) we observe a highly excited network
involving most of the available agents, while for larger systems this
network is much sparser.  However, regardless of the size of the
system, we observe a hierarchical organization reflected by the
multi-modal distribution of activities and connectivities.  When we
increase the size of the system from $12$ to $16$ bits, we observe
that the absolute number of agents participating in the network has
the same order of magnitude. Therefore the number of competent
inactive agents increases as the system size grows, a feature that
might favor the adaptation of large networks to perturbations.

The skewed distributions obtained correspond to the emergence of
structures or self-organization in the network of activated agents,
induced by local interactions between the agents. The system organizes
itself in a hierarchical structure where we find {\it grosso modo\/}
three classes: highly, intermediate an low activated agents. Since we
considered the network of active agents, we do not include among the
classes the non-active one. For the parameters considered in this
paper the Gini coefficient stays always between $0.70-0.80$,
indicating a rather unequal distribution of activities.

In the stationary state we do observe the organization of the system
with a finite number of active agents, without any need of ab initio
considerations about finite population. This means that at this point
there is an equilibrium between agents being deactivated and those
being activated. The model may explain the hierarchy observed in
business networks and social networks. It may be also a good model to
describe the formation of corruption networks. For corruption to take
place, the interaction among the agents in this case cannot be based
on similarities or acquaintances, but rather on complementarity that
generates a dependence among the participants. Diffusion can then be
understood as a way of sharing power in order to protect leading
agents from betrayal. As another application we may consider the
introduction of different types of interactions (other than mutually
reinforcing), which might give rise to multi-modal distributions which
could then be interpreted as the emergence of trophic levels in
ecology.

\paragraph{Acknowledgments}

This work was partially supported by CNPq, CAPES and FAPERJ. We would
like to thank Nestor Caticha and Am\'erico T. Bernardes for their
invaluable suggestions and criticisms.

\bibliography{copelli}

\end{document}